\documentclass{article}
\usepackage{amssymb}
\usepackage{amsfonts}
\usepackage{amsmath}

\input{tcilatex}
\begin{document}

\title{Cosmology from\textbf{\ }Newton--Chern--Simons gravity}
\author{S. Lepe$^{3}\thanks{%
samuel.lepe@pucv.cl}$, G. Rubio$^{2}\thanks{%
gurubio@udec.cl}$ and P. Salgado$^{1,2}\thanks{%
pasalgad@udec.cl}$ \\
$^{1}$Facultad de Ciencias, \ Universidad Arturo Prat de Iquique,\\
Av. Arturo Prat Chac\'{o}n 2120, Iquique, Chile\\
$^{2}$Departamento of F\'{\i}sica, Universidad de Concepci\'{o}n,\\
Casilla 160-C, Concepci\'{o}n, Chile\\
$^{3}$Instituto de F\'{\i}sica, Pontificia Universidad Cat\'{o}lica de
Valpara\'{\i}so,\\
Avenida Brasil 2950, Valpara\'{\i}so, Chile.}
\maketitle

\begin{abstract}
We study a five-dimensional non-relativistic gravity theory whose action is
composed of a gravitational sector and a sector of matter where the
gravitational sector is given by the so called Newton--Chern--Simons gravity
and where the matter sector is described by a perfect fluid. At time to do
cosmology, the obtained field equations shows a close analogy with the
projectable version of the Ho\v{r}ava--Lifshitz theory in $(3+1)$%
-dimensions. Solutions and their asymptotic limits are found. \ In
particular a phantom solution with a future singularity reminiscent of a
Litlle Big Rip future singularity is obtained.
\end{abstract}

\tableofcontents

\section{\textbf{Introduction}}

In Ref. \cite{epjc2014} was studied a five-dimensional
Eins\-tein-Chern-Si\-mons gravity whose action $S=S_{g}+S_{M}$ is composed
of a gravitational sector and a sector of matter, where the gravitational
sector is given by a particular Chern-Simons gravity action \cite{PLB2009}\
instead of the Einstein-Hilbert action and where the matter sector is given
by the so called perfect fluid.

The corresponding Chern-Simons Lagrangian of Ref. \cite{PLB2009} is a
Lagrangian for the so called $\mathfrak{B}$ algebra whose generators $%
\left\{ J_{ab},P_{a},Z_{ab},Z_{a}\right\} $ satisfy the commutation relation
given by in the first equation of Ref. \cite{epjc2014}. This Lagrangian can
be constructed from the one-form gauge connection

\begin{equation}
\boldsymbol{A}=\frac{1}{2}\omega ^{ab}\boldsymbol{J}_{ab}+\frac{1}{l}e^{a}%
\boldsymbol{P}_{a}+\frac{1}{2}k^{ab}\boldsymbol{Z}_{ab}+\frac{1}{l}h^{a}%
\boldsymbol{Z}_{a},
\end{equation}%
and the two-form curvature%
\begin{equation}
\boldsymbol{F}=\frac{1}{2}R^{ab}\boldsymbol{J}_{ab}+\frac{1}{l}T^{a}%
\boldsymbol{P}_{a}+\frac{1}{2}K^{ab}\boldsymbol{Z}_{ab}+\frac{1}{l}H^{a}%
\boldsymbol{Z}_{a},
\end{equation}%
where $T^{a}=\mathrm{D}_{\omega }e^{a},$ $R^{ab}=\mathrm{d}\omega
^{ab}+\omega _{\text{ \ }c}^{a}\omega ^{cb}$, $H^{a}=\mathrm{D}_{\omega
}h^{a}+k_{\text{ \ }b}^{a}e^{b},$ $K^{ab}=\mathrm{D}_{\omega }k^{ab}+\frac{1%
}{l^{2}}e^{a}e^{b}$, are the corresponding "curvatures". \ In fact, using
the extended Cartan's homotopy formula \cite{bzumino,salg3}, and integrating
by parts, we find that the five-dimensional Chern--Simons lagrangian for the 
$\mathfrak{B}$ algebra is given by \cite{epjc2014} 
\begin{eqnarray}
L_{\mathrm{ChS}}^{\left( 5\right) } &=&\alpha _{1}l^{2}\epsilon
_{abcde}R^{ab}R^{cd}e^{e}+\alpha _{3}\epsilon _{abcde}\left( \frac{2}{3}%
R^{ab}e^{c}e^{d}e^{e}+2l^{2}k^{ab}R^{cd}T^{e}+l^{2}R^{ab}R^{cd}h^{e}\right) 
\notag \\
&&+dB_{EChS}^{(4)},  \label{1a}
\end{eqnarray}%
where the surface term $B_{\text{EChS}}^{(4)},$ given by

\begin{align}
B_{\text{EChS}}^{(4)}& =\alpha _{1}l^{2}\epsilon _{abcde}e^{a}\omega
^{bc}\left( \frac{2}{3}\mathrm{d}\omega ^{de}+\frac{1}{2}\omega _{\text{ \ }%
f}^{d}\omega ^{fe}\right)  \notag \\
& \quad +\alpha _{3}\epsilon _{abcde}\left[ l^{2}\left( h^{a}\omega
^{bc}+k^{ab}e^{c}\right) \left( \frac{2}{3}\mathrm{d}\omega ^{de}+\frac{1}{2}%
\omega _{\text{ \ }f}^{d}\omega ^{fe}\right) \right.  \notag \\
& \qquad \qquad \qquad \left. +l^{2}k^{ab}\omega ^{cd}\left( \frac{2}{3}%
\mathrm{d}e^{e}+\frac{1}{2}\omega _{\text{ \ }f}^{d}e^{e}\right) +\frac{1}{6}%
e^{a}e^{b}e^{c}\omega ^{de}\right] .
\end{align}

In the above mentioned reference \cite{epjc2014} and also in Ref.\cite{gomez}
was shown that:

$(i)$ the field equations can be obtained from \ the Lagrangian\ $L=L_{%
\mathrm{ChS}}^{(5)}+\kappa L_{M}$, where $L_{M}=L_{M}(e^{a},h^{a},\omega
^{ab})$ is the matter Lagrangian and\ $\kappa $ is a coupling constant
related to the effective Newton's constant. In fact, the variation of the
lagrangian (\ref{1a}) w.r.t. the dynamical fields vielbein $e^{a}$, spin
connection $\omega ^{ab}$, $h^{a}$ and $k^{ab}$, leads to the following
field equations

\begin{align}
\varepsilon _{abcde}R^{ab}e^{c}e^{d}& =4\kappa _{5}\left( \frac{\delta L_{M}%
}{\delta e^{e}}+\alpha \frac{\delta L_{M}}{\delta h^{e}}\right) ,  \label{9}
\\
\frac{\delta L_{M}}{\delta h^{e}}& =\frac{l^{2}}{8\kappa _{5}}\varepsilon
_{abcde}R^{ab}R^{cd},  \label{10} \\
\varepsilon _{abcde}R^{cd}D_{\omega }h^{e}& =0.  \label{11}
\end{align}%
where we have imposed the conditions \ $T^{a}=0$, $k^{ab}=0$ and $\delta
L_{M}/\delta \omega ^{ab}=0$ and where $\kappa _{5}=\kappa /8\alpha _{3}$
and $\alpha =-\alpha _{1}/\alpha _{3}$. \ Note that the equation (\ref{9})
is analogous to Einstein's equation, where $\delta L_{M}/\delta h^{a}$
correspond to the energy-momentum tensor for the field $h^{a}.$

In the case where the equations (\ref{9}-\ref{11}) satisfy the cosmological
principle and the ordinary matter is negligible compared to the dark energy,
we find that corresponding the FLRW equations \ take the form

\begin{align}
6\left( \frac{{\dot{a}}^{2}+k}{a^{2}}\right) & =\kappa _{5}\alpha \rho
^{(h)},  \label{eqz06} \\
3\left[ \frac{\ddot{a}}{a}+\left( \frac{{\dot{a}}^{2}+k}{a^{2}}\right) %
\right] & =-\kappa _{5}\alpha p^{(h)},  \label{eqz07} \\
{\frac{3l^{2}}{\kappa _{5}}\left( \frac{{\dot{a}}^{2}+k}{a^{2}}\right) ^{2}}%
& =\rho ^{(h)},  \label{eqz08} \\
\frac{3l^{2}}{\kappa _{5}}\frac{\ddot{a}}{a}\left( \frac{{\dot{a}}^{2}+k}{%
a^{2}}\right) & =-p^{(h)},  \label{eqz09} \\
\left( \frac{{\dot{a}}^{2}+k}{a^{2}}\right) \left[ (h-h(0))\frac{\dot{a}}{a}+%
\dot{h}\right] & =0.  \label{eqz10}
\end{align}

These field equations were completely resolved in reference \cite{epjc2014}\
\ for the age of dark energy, where was found that the field $h^{a}$ has a
similar behavior of a cosmological constant.

$\left( ii\right) $The equations (\ref{eqz06}-\ref{eqz10}) have solutions
that describe an accelerated expansion for the three possible cosmological
models of the universe. Namely, spherical expansion $\left( k=1\right) $,
flat expansion $\left( k=0\right) $ and hyperbolic expansion $\left(
k=-1\right) $ when the constant $\alpha $ is greater than zero. This mean
that the FRW--Einstein--Chern-Simons field equations have as a of their
solutions an universe in accelerated expansion. This result allow us to
conjeture that this solutions are compatible with the era of Dark Energy and
that the energy-momentum tensor for the field $h^{a}$ corresponds to a form
of positive cosmological constant.

In summary in Refs. \cite{epjc2014,gomez} were studied the implications of
replacing in the action $S=S_{g}+S_{M}$ the Einstein--Hilbert action by the
Einstein--Chern--Simons action on the cosmological evolution for a
Friedmann--Lema\^{\i}tre--Robertson--Walker metric (FLRW). In the case that
the matter action $S_{M}$ is the action for a perfect fluid, was found that
the FRW--Einstein--Chern--Simons field equations have solutions that
describe an accelerated expansion for the three possible cosmological models
of the universe.

On the other hand, in Ref. \cite{salg1} was found that the non-relativistic
limit of Einstein--Chern--Simons gravity action is given by the so called
Newton--Chern--Simons gravity action. This action is invariant under the so
called non-relativistic algebra $\mathcal{G}\mathfrak{B}_{_{5}},$ which can
be obtained as the non-relativistic limit of the generalized Poincar\'{e}
algebra $\mathfrak{B}_{_{5}}.$

One of the purpose of this article is to find a non-relativistic limit of
the results found in references \cite{epjc2014,gomez}, i.e., some
cosmological solutions for the field equations which can be obtained from
the Newton--Chern--Simons action studied in Ref. \cite{salg1}. \ 

This paper is organized as follows: In Section $2$ we obtain the field
equations for the Lagrangian $L=L_{\mathrm{ChS}}^{(5)}+\kappa L_{M}$, where $%
L_{\mathrm{ChS}}^{(5)}$ is the Newton--Chern--Simons Lagrangian and $L_{M}$
is the corresponding matter Lagrangian. These field equations correspond to
the non-relativistic limit of the field equations studied in Refs. \cite%
{epjc2014,gomez}.\ In Section $3$ we find the field equations for a
Newton--Chern--Simons cosmology. In Section $4$ it is shown that the
Newton--Chern--Simons cosmology is a sort of analogue of the projectable
version of the Ho\v{r}ava--Lifshitz theory in $(3+1)$-dimensions, although
one of the terms is not present. Solutions and their asymptotic limits are
found, which show interesting properties. In particular a phantom solution
with a future singularity reminiscent of a Litlle Big Rip future singularity
is obtained. Finally, a brief revision of the adiabaticity in the cosmic
evolution is made.

\section{\textbf{Newton--Chern--Simons gravity}}

In this section we will make a brief review of the so-called
Newton--Chern--Simons gravity. \ The non-relativistic algebra $\mathcal{G}%
\mathfrak{B}_{_{5}}$ has the following commutation relation \cite{salg1}
(see also \cite{salg2}),

\begin{align}
\lbrack J_{ij},J_{kl}]& =\eta _{kj}J_{il}+\eta _{lj}J_{ki}-\eta
_{ki}J_{jl}-\eta _{li}J_{kj},  \notag \\
\lbrack J_{ij},K_{k}]& =\eta _{jk}K_{i}-\eta _{ik}K_{j},\text{ \ \ }%
[K_{i},P_{j}]=-\delta _{ij}M,  \notag \\
\lbrack J_{ij},P_{k}]& =\eta _{jk}P_{i}-\eta _{ik}P_{j},\text{ \ }%
[K_{i},H]=-P_{i},  \notag \\
\lbrack J_{ij},Z_{kl}]& =\eta _{kj}Z_{il}+\eta _{lj}Z_{ki}-\eta
_{ki}Z_{jl}-\eta _{li}Z_{kj},  \notag \\
\lbrack J_{ij},Z_{k0}]& =\eta _{jk}Z_{i0}-\eta _{ik}Z_{j0},\text{ \ \ }%
[K_{i},Z_{j}]=-\delta _{ij}N,  \notag \\
\lbrack Z_{ij},K_{k}]& =\eta _{jk}Z_{i0}-\eta _{ik}Z_{j0},\text{ \ }%
[K_{i},Z_{0}]=-Z_{i},  \notag \\
\lbrack J_{ij},Z_{k}]& =\eta _{jk}Z_{i}-\eta _{ik}Z_{j},\text{ \ \ }%
[Z_{i0},P_{j}]=-\delta _{ij}N,  \notag \\
\lbrack Z_{ij},P_{k}]& =\eta _{jk}Z_{i}-\eta _{ik}Z_{j},\text{ \ }%
[Z_{i0},H]=-Z_{i},  \notag \\
\lbrack P_{i},H]& =\text{ }Z_{i0}.  \label{gb5}
\end{align}%
The one-form gauge connetion $A$ valued in the $\mathcal{G}\mathfrak{B}%
_{_{5}}$ algebra is given by 
\begin{align}
A& =\frac{c}{l}\tau H+\frac{1}{l}e^{i}P_{i}+\frac{c}{l}\hat{\tau}Z_{0}+\frac{%
1}{l}h^{i}Z_{i}+\frac{1}{cl}mM+\frac{1}{cl}nN  \notag \\
& +\frac{1}{c}\omega ^{i}K_{i}+\frac{1}{c}k^{i}Z_{i0}+\frac{1}{2}\omega
^{ij}J_{ij}+\frac{1}{2}k^{ij}Z_{ij},
\end{align}%
where $l$ and $c$ are parameters of dimensions of length and velocity
respectively. \ The corresponding 2-form curvature $F=dA+AA$ is then given
by \cite{salg1}%
\begin{align}
F& =\frac{c}{l}R(H)H+\frac{1}{l}R^{i}(P_{i})P_{i}+\frac{c}{l}R(Z_{0})Z_{0}+%
\frac{1}{l}R^{i}(Z_{i})Z_{i}+\frac{1}{cl}R(M)M  \notag \\
& +\frac{1}{cl}R(N)N+\frac{1}{c}R^{i}\left( K_{i}\right) K_{i}+\frac{1}{c}%
R^{i}\left( Z_{i0}\right) Z_{i0}+\frac{1}{2}R^{ij}\left( J_{ij}\right)
J_{ij}+\frac{1}{2}R^{ij}\left( Z_{ij}\right) Z_{ij},
\end{align}%
\newline
where 
\begin{align}
R(H)& =d\tau ,\text{ }R^{i}(P_{i})=T^{i}-\omega ^{i}\tau ,  \notag \\
R(Z_{0})& =d\hat{\tau},\text{ \ }R(M)=dm-\omega ^{i}e_{i},  \notag \\
R^{i}(Z_{i})& =Dh^{i}-\omega ^{i}\hat{\tau}-k^{i}\tau +k_{\,\,j}^{i}e^{j}, 
\notag \\
R(N)& =dn-\omega ^{i}h_{i}-k^{i}e_{i},  \notag \\
R^{i}(Z_{i0})& =Dk^{i}+\nu ^{2}e^{i}\tau +k_{\,\,j}^{i}\omega ^{j},\text{ \ }%
R^{i}(K_{i})=D\omega ^{i},  \notag \\
R^{ij}(J_{ij})& =R^{ij},\text{ \ \ }R^{ij}(Z_{ij})=Dk^{ij},
\end{align}%
with $\nu =c/l$, $T^{i}=de^{i}+\omega ^{ij}e_{j}$ and $R^{ij}=d\omega
^{ij}+\omega _{\,\,k}^{i}\omega ^{kj}$.\newline

From the gauge connection transformation for $A$, $\delta A=d\Lambda +\left[
A,\Lambda \right] $, with%
\begin{align}
\Lambda & =\frac{v}{l}\zeta ^{0}H+\frac{1}{l}\zeta ^{i}P_{i}+\frac{v}{l}\rho
^{0}Z_{0}+\frac{1}{l}\rho ^{i}Z_{i}+\frac{1}{vl}\sigma M+\frac{1}{vl}\gamma N
\notag \\
& +\frac{1}{v}\lambda ^{i}K_{i}+\frac{1}{v}\chi ^{i}Z_{i0}+\frac{1}{2}%
\lambda ^{ij}J_{ij}+\frac{1}{2}\chi ^{ij}Z_{ij},
\end{align}%
it is direct to find the variations of the diferent gauge fields \cite{salg1}
\begin{align}
\delta \tau & =d\zeta ^{0}\text{, \ }\delta e^{i}=D\zeta ^{i}-\omega
^{i}\zeta ^{0}-\lambda ^{ij}e_{j}+\tau \lambda ^{i},  \notag \\
\delta h^{i}& =D\rho ^{i}-\omega ^{i}\rho ^{0}-\lambda
^{ij}h_{j}+h^{0}\lambda ^{i}+k^{ij}\zeta _{j}-k^{i}\zeta ^{0}-\chi
^{ij}e_{j}+\tau \chi ^{i},  \notag \\
\delta m& =d\sigma -\omega ^{i}\zeta _{i}+e^{i}\lambda _{i}\text{, \ }\delta
\omega ^{i}=D\lambda ^{i}-\lambda ^{ij}\omega _{j},  \notag \\
\delta n& =d\gamma -k^{i}\zeta _{i}+h^{i}\lambda _{i}-\omega ^{i}\rho
_{i}+e^{i}\chi _{i}\text{,\ }\delta h^{0}=d\rho ^{0},  \notag \\
\delta k^{i}& =D\chi ^{i}-\lambda ^{ij}k_{j}-\chi ^{ij}\omega
_{j}+k^{ij}\lambda _{j}+e^{i}\zeta ^{0}-\zeta ^{i}\tau ,\text{ \ }\delta
\omega ^{ij}=D\lambda ^{ij},  \notag \\
\delta k^{ij}& =D\chi ^{ij}+k_{\,\,k}^{i}\lambda ^{kj}+k_{\,\,k}^{j}\lambda
^{ik},  \label{campos}
\end{align}%
where the derivative $D$ is covariant with respect to the $J$%
-transformations.

From (\ref{campos}) we can see that only the gauge fields $e_{\mu }^{\,\,i}$%
, $\tau _{\mu }$, $m_{\mu }$, $h_{\mu }^{\,\,i}$, $h_{\mu }^{0}$ and $n_{\mu
}$ transform under $P$ and $H$ transformations. These are the fields that
should remain independent, while the remaining fields will be dependent upon
the aforementioned fields. This can be archieved with the following
constraints%
\begin{align}
R(H)& =d\tau =0,\text{ \ }R^{i}(P_{i})=T^{i}-\omega ^{i}\tau =0,  \notag \\
R(M)& =dm-\omega ^{i}e_{i}=0,\text{ \ }R(Z_{0})=dh^{0}=0,  \notag \\
R^{i}(Z_{i})& =Dh^{i}-\omega ^{i}h^{0}-k^{i}\tau +k_{j}^{i}e^{j}=0,  \notag
\\
R(N)& =dn-\omega ^{i}h_{i}-k^{i}e_{i}=0.  \label{constraints}
\end{align}

Using the subspaces separation method introduced in Ref. \cite{salg3}, was
found that, except for surface terms, the so called Newton--Chern--Simons
lagrangian is given by

\begin{align}
L_{\mathrm{NRChS}}& =\alpha _{1}\varepsilon _{ijkl}\left(
-2R^{ij}T^{k}\omega ^{l}-\frac{4}{3}R^{ij}\omega ^{k}\omega ^{l}\tau
+2R^{ij}D\omega ^{k}e^{l}-R^{ij}R^{kl}m\right)  \notag \\
& \,\,\,\,+\alpha _{3}\varepsilon _{ijkl}\left( \frac{4}{3}\nu
^{2}R^{ij}e^{k}e^{l}\tau -2R^{ij}Dh^{k}\omega ^{l}-\frac{4}{3}%
R^{ij}k^{k}\omega ^{l}\tau -\frac{4}{3}R^{ij}\omega ^{k}\omega ^{l}\hat{\tau}%
\right.  \notag \\
& \left. +2R^{ij}D\omega ^{k}h^{l}-\frac{4}{3}Dk^{ij}T^{k}\omega
^{l}-Dk^{ij}\omega ^{k}\omega ^{l}\tau -R^{ij}k^{kl}dm-\frac{2}{3}%
R^{ij}k^{kl}e^{m}\omega _{m}\right.  \notag \\
& \,\,\,\,\left. -\frac{2}{3}R^{ij}\omega _{\,m}^{k}k^{ml}m-\frac{4}{3}%
k^{ij}T^{k}D\omega ^{l}-k^{ij}D\omega ^{k}\omega ^{l}\tau -2R^{ij}T^{k}k^{l}-%
\frac{4}{3}R^{ij}\omega ^{k}k^{l}\tau \right.  \notag \\
& \,\,\,\,\left. +\frac{2}{3}R^{ij}k^{km}\omega _{m}e^{l}+\frac{2}{3}\omega
_{\,m}^{i}k^{jm}D\omega ^{k}e^{l}-R^{ij}R^{kl}n-2R^{ij}\omega
^{km}k_{m}e^{l}\right) ,  \label{lagrangeanob}
\end{align}%
where $\nu ,\alpha _{1},\alpha _{3}$ are parameters of the theory and $%
\kappa $ is a constants (for detail see \cite{epjc2014,gomez,salg1}). \ 

In the next section we will consider obtaining the equations of motion
associated with the action whose Lagrangian is given by the eq. (\ref%
{lagrangeanob})

\section{\textbf{Newton--Chern--Simons field equations}}

In presence of matter, the complete Lagrangian of the theory is 
\begin{equation}
L=\kappa L_{M}+L_{\mathrm{NRChS}}  \label{lagran1}
\end{equation}%
where $L_{\mathrm{NRChS}}$ is the Newton--Chern--Simons lagrangian given in (%
\ref{lagrangeanob}) and $L_{M}$ is the corresponding matter Lagrangian.

The field equations obtained from the action (\ref{lagran1}) are given by

\begin{eqnarray}
\varepsilon _{ijkl}\left( -\frac{4}{3}\alpha _{1}R^{ij}\omega ^{k}\omega
^{l}+\frac{4}{3}\alpha _{3}\nu ^{2}R^{ij}e^{k}e^{l}\right) &=&\kappa \frac{%
\delta L_{M}}{\delta \tau },  \label{1} \\
\frac{4}{3}\alpha _{3}\varepsilon _{ijkl}R^{ij}\omega ^{k}\omega ^{l}
&=&-\kappa \frac{\delta L_{M}}{\delta \hat{\tau}},  \label{2} \\
2\varepsilon _{ijkl}\left( \alpha _{1}R^{ij}D\omega ^{k}-\frac{4}{3}\alpha
_{3}\nu ^{2}R^{ij}e^{k}\tau \right) &=&\kappa \frac{\delta L_{M}}{\delta
e^{l}},  \label{3} \\
2\alpha _{3}\varepsilon _{ijkl}R^{ij}D\omega ^{k} &=&\kappa \frac{\delta
L_{M}}{\delta h^{l}},  \label{4} \\
a_{1}\varepsilon _{ijkl}R^{ij}R^{kl} &=&-\kappa \frac{\delta L_{M}}{\delta m}%
,  \label{5} \\
a_{3}\varepsilon _{ijkl}R^{ij}R^{kl} &=&-\kappa \frac{\delta L_{M}}{\delta n}%
,  \label{6}
\end{eqnarray}%
\begin{equation}
4\varepsilon _{ijkl}\left( \frac{2}{3}\alpha _{1}R^{ij}\omega ^{k}\tau
-\alpha _{1}R^{ij}T^{k}+\frac{2}{3}\alpha _{3}R^{ij}\omega ^{k}\hat{\tau}%
-\alpha _{3}R^{ij}Dh^{k}\right) =\kappa \frac{\delta L_{M}}{\delta \omega
^{l}},  \label{7}
\end{equation}%
\begin{eqnarray}
&&\varepsilon _{ijkl}\left( -2\alpha _{1}R^{km}e_{m}\omega ^{l}-4\alpha
_{1}T^{k}D\omega ^{l}-\frac{4}{3}\alpha _{1}\omega ^{k}\omega ^{l}d\tau -%
\frac{8}{3}\alpha _{1}D\omega ^{k}\omega ^{l}\tau +2\alpha _{1}R^{km}\omega
_{m}e^{l}\right.  \notag \\
&&-2\alpha _{1}R^{kl}dm+\frac{8}{3}\nu ^{2}\alpha _{3}T^{k}e^{l}\tau +\frac{4%
}{3}\alpha _{3}e^{k}e^{l}d\tau -2\alpha _{3}R^{km}h_{m}\omega ^{l}-4\alpha
_{3}Dh^{k}D\omega ^{l}  \notag \\
&&\left. -\frac{4}{3}\alpha _{3}\omega ^{k}\omega ^{l}d\hat{\tau}-\frac{8}{3}%
\alpha _{3}D\omega ^{k}\omega ^{l}\hat{\tau}+2\alpha _{3}R^{km}\omega
_{m}h^{l}-\alpha _{3}R^{kl}dn\right) =\kappa \frac{\delta L_{M}}{\delta
\omega ^{ij}},  \label{8}
\end{eqnarray}%
where we have imposed the $k^{ij}=k^{i}=0$ conditions, and used

\begin{equation*}
T_{i}=\ast \left( \frac{\delta L_{M}}{\delta e^{i}}\right) ,\text{ \ }%
T_{0}=\ast \left( \frac{\delta L_{M}}{\delta \tau }\right) .
\end{equation*}

From (\ref{constraints}) and Bianchi identities we find 
\begin{equation}
DT^{i}=D\omega ^{i}\tau \,\,,\,\,\,\,R^{ij}e_{j}=D\omega ^{i}\tau \,,\text{
\ }R_{[\lambda \mu }^{\,\,\,\,\,\,\,\,ij}e_{\nu ]j}^{\,}=(D_{[\lambda
}\omega _{\mu })^{i}\tau _{\nu ]},  \label{bianchi}
\end{equation}

\begin{equation}
e_{[\lambda i}(D_{\mu }\omega _{\nu ]})^{i}=0.  \label{bianchi2}
\end{equation}%
For simplicity we will assume that the torsion vanishes. In this case $%
\delta L_{M}/\delta \omega ^{ij}=0$. Using the constraint (\ref{constraints}%
) we find that (\ref{7}) takes the form

\begin{align}
4\varepsilon _{ijkl}\left( \frac{\alpha _{1}}{3}R^{ij}\omega ^{k}\tau +\frac{%
\alpha _{3}}{3}R^{ij}\omega ^{k}\hat{\tau}\right) & =0,  \notag \\
\hat{\tau}& =-\frac{\alpha }{3}\tau .  \label{tao}
\end{align}%
Introducing (\ref{bianchi}),(\ref{bianchi2}) in (\ref{8}) we have 
\begin{eqnarray}
&&\varepsilon _{ijkl}\left( -\frac{10\alpha _{1}}{3}D\omega ^{k}\omega
^{l}\tau +2\alpha _{1}R^{km}\omega _{m}e^{l}-\frac{8}{3}\nu ^{2}\alpha
_{3}\omega ^{k}\tau e^{l}\tau -\alpha _{1}R^{kl}dm\right.  \notag \\
&&\left. +2\alpha 3R^{km}\omega _{m}h^{l}-\frac{10}{3}\alpha _{3}D\omega
^{k}\omega ^{l}\hat{\tau}-\alpha _{3}R^{kl}dn\right)  \notag \\
&=&0.
\end{eqnarray}%
Since $\tau ^{2}=0$ we can write 
\begin{equation}
\varepsilon _{ijkl}\left( -\frac{13\alpha _{1}}{3}D\omega ^{k}\omega
^{l}\tau -\frac{13}{3}\alpha _{3}D\omega ^{k}\omega ^{l}\hat{\tau}\right) =0,
\end{equation}%
which means that this equation is satisfied identically and therefore the
space is a flat manifold as can be seen from the equations (\ref{5}), (\ref%
{6}).

Introducing eqs. (\ref{2}) in (\ref{1}) and (\ref{4}) in (\ref{3}), we
obtain 
\begin{eqnarray}
\varepsilon _{ijkl}R^{ij}e^{k}e^{l} &=&\frac{6}{\nu ^{2}}\left( k_{1}\frac{%
\delta L_{M}}{\delta \tau }-\alpha k_{2}\frac{\delta L_{M}}{\delta \hat{\tau}%
}\right) ,  \notag \\
\varepsilon _{ijkl}R^{ij}e^{k}\tau &=&\frac{3}{\nu ^{2}}\left( k_{1}\frac{%
\delta L_{M}}{\delta e^{l}}-\alpha \frac{\delta L_{M}}{\delta h^{l}}\right) .
\end{eqnarray}%
Taking into account that

\begin{eqnarray}
\ast (T_{0})\delta \tau &=&\det (g)\delta _{\delta }^{\sigma }T_{\text{ \ }%
0}^{0}\text{\ }\delta \tau _{\text{ \ }\sigma }^{\delta }dx^{5},
\label{coord1} \\
\varepsilon _{ijkl0}R^{ij}e^{k}e^{l}\delta \tau &=&2\det (g)\left( \delta
_{\delta }^{\sigma }R-2R_{\text{ \ }\delta }^{\sigma }\right) \delta \tau _{%
\text{ \ }\sigma }^{\delta }dx^{5},  \label{coord2} \\
\varepsilon _{ijkl0}R^{ij}e^{k}\tau \delta e^{l} &=&-2\det (g)\left( \delta
_{\delta }^{\sigma }R-2R_{\text{ \ }\delta }^{\sigma }\right) \delta e_{%
\text{ \ }\sigma }^{\delta }dx^{5},  \label{coord3}
\end{eqnarray}%
and using $T_{\text{ \ }00}^{(h)}=\rho ^{(h)}$, $T_{\text{ \ }%
ii}^{(h)}=4p^{(h)}/c^{2}$, we find (with $R=0$)

\begin{equation}
R_{00}=\frac{3}{2\nu ^{2}}\left( k_{1}\rho ^{(e)}-\alpha k_{2}\rho
^{(h)}\right) ,  \label{r00}
\end{equation}%
\begin{equation}
R_{00}=\frac{3}{c^{2}\nu ^{2}}\left( k_{1}p^{(e)}-\alpha k_{2}p^{(h)}\right)
,  \label{r002}
\end{equation}%
where (\ref{r00}) coincides with the results found in \cite{salg1}

\begin{equation}
\nabla ^{2}\phi =\frac{3}{2\nu ^{2}}(k_{1}\rho ^{(e)}-\alpha k_{2}\rho
^{(h)}),
\end{equation}%
with $\nu =c/l,$ $\beta _{1}=\beta _{2}=\kappa $, $k_{1}=\kappa /8\alpha
_{3}=8\pi G_{5}$, $k_{2}=\kappa /24\alpha _{3},$ $\alpha =3\alpha
_{1}/\alpha _{3},$ $k_{1}=3k_{2}.$ From (\ref{r00}), (\ref{r002}) we have, 
\begin{eqnarray}
2k_{1}p^{(e)}-2\alpha k_{2}p^{(h)} &=&\left( k_{1}\rho ^{(e)}-\alpha
k_{2}\rho ^{(h)}\right) c^{2},  \notag \\
2p^{(e)}-\frac{2\alpha }{3}p^{(h)} &=&\left( \rho ^{(e)}-\frac{\alpha }{3}%
\rho ^{(h)}\right) c^{2}.
\end{eqnarray}%
Defining a density and an effective pressure as%
\begin{eqnarray}
p &=&\frac{p^{(e)}}{2}-\frac{\alpha }{6}p^{(h)},  \notag \\
\rho &=&\frac{\rho ^{(e)}}{2}-\frac{\alpha }{6}\rho ^{(h)},
\end{eqnarray}%
we find%
\begin{equation}
p=\frac{\rho c^{2}}{2},  \label{estad}
\end{equation}%
and from (\ref{tao}), we have 
\begin{equation}
\rho ^{(h)}=-\frac{3}{\alpha }\rho ^{(e)},\text{\ }  \label{densidad}
\end{equation}%
From (\ref{r00}), (\ref{r002}) we can see%
\begin{eqnarray}
R_{00} &=&\frac{3}{4\nu ^{2}}\left( k_{1}\rho ^{(e)}-\alpha k_{2}\rho
^{(h)}+2k_{1}\frac{p^{(e)}}{c^{2}}-2\alpha k_{2}\frac{p^{(h)}}{c^{2}}\right)
,  \notag \\
R_{00} &=&\nabla ^{2}\phi =\frac{3k_{1}}{2\nu ^{2}}\left( \rho +\frac{2p}{%
c^{2}}\right) .  \label{poisson2}
\end{eqnarray}

On the other hand, the interaction between the fluids is described by the
following state equations

\begin{eqnarray}
p^{(e)} &=&\omega ^{(e)}\rho ^{(e)}c^{2},  \notag \\
p^{(h)} &=&\omega ^{(h)}\rho ^{(h)}c^{2}=-\frac{3}{\alpha }\omega ^{(h)}\rho
^{(e)}c^{2},
\end{eqnarray}%
\begin{eqnarray}
2k_{1}\omega ^{(e)}\rho ^{(e)}-2\alpha k_{2}\omega ^{(h)}\rho ^{(h)}
&=&k_{1}\rho ^{(e)}-\alpha k_{2}\rho ^{(h)},  \notag \\
2\left( k_{1}\omega ^{(e)}+3k_{2}\omega ^{(h)}\right) \rho ^{(e)} &=&\left(
k_{1}+3k_{2}\right) \rho ^{(e)},
\end{eqnarray}%
\begin{eqnarray}
\omega ^{(h)} &=&\frac{\left( k_{1}+3k_{2}\right) }{6k_{2}}-\frac{k_{1}}{%
3k_{2}}\omega ^{(e)},  \notag \\
&=&1-\omega ^{(e)}.  \label{rel}
\end{eqnarray}

In the next section we will study a possible non-relativistic version of the
results obtained in Ref. \cite{epjc2014}.

\section{\textbf{Newton--Chern--Simons cosmology}}

Following the formalism used in \cite{2}, we denote with $(t,x^{i})$, the
local coordinates where $i=1,2,3,4$ and $\tau =dx^{0},$ $h=\delta
^{ij}\partial _{i}\otimes \partial _{j}$ are the temporal and spatial metric
respectively. The matter is modeled as an ideal fluid with velocity $u$,
which is a timelike unit vector. The vorticity\ $\Omega ^{\alpha \beta }$
and the (rate of) strain $\Theta ^{\alpha \beta }$ relative to a timelike
unit vector field $V$, where $\tau (V)=1$, i.e., $\tau _{a}=g_{\alpha \beta
}V^{\beta }$, are given by

\begin{eqnarray}
\Omega ^{\alpha \beta } &=&\frac{1}{2}(u_{;\lambda }^{\alpha }h^{\lambda
\beta }-u_{;\lambda }^{\beta }h^{\lambda \alpha }),  \notag \\
\Theta ^{\alpha \beta } &=&\frac{1}{2}(u_{;\lambda }^{\alpha }h^{\lambda
\beta }+u_{;\lambda }^{\beta }h^{\lambda \alpha }).
\end{eqnarray}%
The expansion rate and the (rate of) shear is the trace-free part of the
strain are given by 
\begin{equation}
\theta =h^{\alpha \beta }\Theta _{\alpha \beta },\text{ \ \ }\sigma =\Theta -%
\frac{1}{4}\theta h
\end{equation}%
respectively.\ \ 

It is posible to show that $\theta =u_{;\sigma }^{\sigma }$ and that the
covariant derivative of the velocity can be decomposed as \cite{2} 
\begin{equation}
h_{\alpha \lambda }u_{;\beta }^{\lambda }=\Theta _{\alpha \beta }+\Omega
_{\alpha \beta }+h_{\alpha \rho }V^{\lambda }u_{;\lambda }^{\rho }g_{\beta
\sigma }V^{\sigma },
\end{equation}%
and with the help of this last equation we can obtain the so called
Raychaudhuri equation in the Newton--Chern--Simons gravity. Following Ref. 
\cite{3} we start from the known identity (see also \cite{11}) 
\begin{eqnarray}
u_{;\beta ;\gamma }^{\alpha }-u_{;\gamma ;\beta }^{\alpha } &=&R_{\sigma
\gamma \beta }^{\alpha }u^{\sigma },  \notag \\
u^{\beta }u_{;\alpha ;\beta }^{\alpha } &=&\left( u^{\beta }u_{;\beta
}^{\alpha }\right) _{;\alpha }-u_{;\alpha }^{\beta }u_{;\beta }^{\alpha
}-R_{\alpha \beta }u^{\alpha }u^{\beta },
\end{eqnarray}%
where the two first terms on the right are given by \cite{2}

\begin{eqnarray}
\left( u^{\beta }u_{;\beta }^{\alpha }\right) _{;\alpha } &=&\func{div}%
(\nabla _{u}u),  \notag \\
u_{;\alpha }^{\beta }u_{;\beta }^{\alpha } &=&h^{\rho \alpha }h^{\sigma
\beta }(\Theta _{\rho \beta }\Theta _{\sigma \alpha }+\Omega _{\rho \beta
}\Omega _{\sigma \alpha }),
\end{eqnarray}%
and the last terms on the right is given by 
\begin{equation}
R_{\alpha \beta }u^{\alpha }u^{\beta }=\frac{3k_{1}}{2\nu ^{2}}\left( \rho +%
\frac{2p}{c^{2}}\right) ,
\end{equation}%
where we have used (\ref{poisson2}) together to the equations 
\begin{eqnarray}
R_{\alpha \beta } &=&\frac{3k_{1}}{2\nu ^{2}}\left( \rho +\frac{2p}{c^{2}}%
\right) \tau _{\alpha }\tau _{\beta },  \notag \\
\tau _{\alpha }\tau _{\beta } &=&g_{\alpha \sigma }g_{\beta \rho }u^{\sigma
}u^{\rho }.  \label{r1}
\end{eqnarray}%
These results allow us to find the five dimensional Raychaudhuri equation
for the Newton--Chern--Simons gravity 
\begin{equation}
\func{div}(\nabla _{u}u)=\nabla _{u}\theta +\frac{1}{4}\theta ^{2}+\sigma
^{\alpha \beta }\sigma _{\alpha \beta }-\Omega ^{\alpha \beta }\Omega
_{\alpha \beta }+\frac{3k_{1}}{2\nu ^{2}}\left( \rho +\frac{2p}{c^{2}}%
\right) .  \label{ray1}
\end{equation}

\subsection{\textbf{FLRW background}}

In this section we study \ the non-relativistic FLRW equations in the
context of the Newton-Chern-Simons gravity.

The calculation of the Ricci tensor from its definition leads to the
following result

\begin{eqnarray}
R_{00} &=&-\frac{1}{2}(h^{ij}\dot{h}_{ij})_{,0}-\frac{1}{4}h^{ij}\dot{h}%
_{jk}h^{kl}\dot{h}_{li}+2h^{ij}\kappa _{0j,i}+\kappa ^{ij}\kappa _{ij}, 
\notag \\
&=&\frac{3k_{1}}{2\nu ^{2}}\left( \rho +\frac{2p}{c^{2}}\right) ,  \notag \\
R_{0i} &=&h^{jk}\kappa _{ik,j}=0,  \notag \\
R_{ij} &=&0.  \label{ray2}
\end{eqnarray}%
The first equation is equivalent to the Raychaudhuri equation (\ref{ray1})
for $u=V,$ while the second equation is equivalent to $\Omega _{\text{ \ }%
,j}^{ij}=0$ for $u=V$, since for any $u$ 
\begin{equation}
\Omega _{\alpha \beta }=\frac{1}{2}\left(
h_{ac}u_{,b}^{c}-h_{bc}u_{,a}^{c}-2\kappa _{ab}\right) .
\end{equation}%
For the other kinematical quantities we find%
\begin{eqnarray}
\Theta _{ab} &=&\frac{1}{2}\left( h_{ac}u_{,b}^{c}-h_{bc}u_{,a}^{c}+\dot{h}%
_{ab}\right) ,  \label{cant1} \\
\theta &=&u_{,a}^{a}+\frac{1}{2}h^{ab}\dot{h}_{ab},  \label{cant2} \\
\func{div}(\nabla _{u}u) &=&\dot{u}_{,a}^{a}+2h^{ab}\kappa
_{0a,b}+2h^{bc}\kappa _{ac}u_{,b}^{a}  \notag \\
&&+h^{bc}u_{,b}^{a}\dot{h}_{ac}+u^{a}u_{,ab}^{b}+u_{,b}^{a}u_{,a}^{b}.
\label{cant3}
\end{eqnarray}%
For an ideal fluid with pressure, the continuity equation and the Euler
equation are respectively given by \cite{10}, 
\begin{equation}
\dot{\rho}+\left[ \left( \rho +\frac{p}{c^{2}}\right) u^{i}\right] _{,i}+%
\frac{1}{2}h^{ij}\dot{h}_{ij}\left( \rho +\frac{p}{c^{2}}\right) =0,
\label{cont}
\end{equation}%
and%
\begin{equation}
\dot{u}^{i}+u^{j}u_{,j}^{i}+2h^{ij}\kappa _{0j}+2u^{j}\left( \frac{1}{2}%
h^{ik}\dot{h}_{kj}+h^{ik}\kappa _{jk}\right) +\left( \rho +\frac{p}{c^{2}}%
\right) ^{-1}h^{ij}p_{,j}=0.  \label{euler}
\end{equation}%
When $\kappa _{ij}=0$, we find that the equations (\ref{cont}), (\ref{euler}%
) and (\ref{ray2}) take the form 
\begin{eqnarray}
\dot{\rho}+\left[ \left( \rho +\frac{p}{c^{2}}\right) u^{i}\right] _{,i}
&=&0,  \notag \\
\dot{u}^{i}+u^{j}u_{,j}^{i} &=&-\left( \rho +\frac{p}{c^{2}}\right)
^{-1}p_{,i}+g^{i},  \notag \\
-g_{,i}^{i} &=&\frac{3k_{1}}{2\nu ^{2}}\left( \rho +\frac{2p}{c^{2}}\right) ,
\end{eqnarray}%
where $g^{i}=-2\kappa _{0i}$. \ So that we have arrived to the equations for
Newton--Chern--Simons gravity coupled to an ideal fluid.

If now we assume that $\rho $ and $p$ are only functions of $t$
(homogeneity), then the Euler equation implies 
\begin{equation}
\nabla _{u}u=-\left( \rho +\frac{p}{c^{2}}\right) ^{-1}\func{div}(ph)=0,
\end{equation}%
and the continuity equation (\ref{cont}) shows that $u_{,i}^{i}$ depends
only of the time $t$. These results leads to the following simplifications
of the equation (\ref{ray1}) 
\begin{equation}
\dot{\theta}+\frac{1}{4}\theta ^{2}+\sigma ^{ab}\sigma _{ab}-\Omega
^{ab}\Omega _{ab}+\frac{3k_{1}}{2\nu ^{2}}\left( \rho +\frac{2p}{c^{2}}%
\right) =0.  \label{theta}
\end{equation}%
Since we have used the fact that $\theta $ is a function that depends only
on time, we have that (\ref{cant2}) and (\ref{cont}) imply 
\begin{equation}
\dot{\rho}+\theta \left( \rho +\frac{p}{c^{2}}\right) =0,  \label{ro}
\end{equation}%
Let us now consider a homogeneous and isotropic flat-FLRW background in the
context of Newton--Chern--Simons gravity. This model is found using the
following Ansatz 
\begin{equation}
V=u,\text{ \ }h_{ij}=a^{2}(t)\delta _{ij},\text{ }\Omega =0,
\end{equation}%
which leads to 
\begin{equation}
\theta =4\frac{\dot{a}}{a},\text{ \ }\sigma _{ab}=0,\text{ \ }\dot{\theta}%
=4\left( \frac{\ddot{a}a-\dot{a}^{2}}{a^{2}}\right) .
\end{equation}%
Here, $a$ is the cosmic scale factor. Introducing these results in the
equations (\ref{ro}) and (\ref{theta}) we obtain

\begin{eqnarray}
\dot{\theta}+\frac{1}{4}\theta ^{2} &=&-\frac{3k_{1}}{2\nu ^{2}}\left( \rho +%
\frac{2p}{c^{2}}\right) ,  \notag \\
\frac{\ddot{a}}{a} &=&-\frac{3k_{1}}{8\nu ^{2}}\left( \rho +\frac{2p}{c^{2}}%
\right) .  \label{theta2}
\end{eqnarray}

In the following Section we will use the equations (\ref{theta2}) to
visualize the cosmologies that can be derived from the present
five-dimensional scheme

\subsection{\textbf{Cosmological solutions}}

From now on we use units $k_{1}=8\pi G_{5}=1=c$ and $\nu =1/l$. The
equations (\ref{theta2}) are the conservation equation and the equation for
the acceleration, respectively, where $p$ is the pressure, $\rho $ the
energy density, $\theta =4H$, $H=\dot{a}/a$ is the Hubble parameter and $a$
is the cosmic scale factor. We immediately visualize the absence of the
Friedmann constraint. This situation is analogous to what happens in the
projectable version of the Ho\v{r}ava--Lifshitz theory in (3+1)-dimensions 
\cite{[1]}. From equations (\ref{theta2}) it is possible to obtain the first
integral

\begin{equation}
\frac{8}{3}\nu ^{2}H^{2}=\rho +\frac{C_{0}}{a^{2}},  \label{3'}
\end{equation}%
where $C_{0}$ is an integration constant. The term $C_{0}/a^{2}$ is not dark
matter in the usual sense, but gravitationally behaves like a fluid whose
pressure is $p=-\left( 1/2\right) \rho $ which, as we shall see, corresponds
to an evolutionary scheme with zero acceleration, which is a Milne universe.
In General Relativity in (3+1)-dimensions, dark matter corresponds to $\rho
\left( a\right) =\rho \left( a_{0}\right) \left( a_{0}/a\right) ^{3}$. A
term of this form is present in the Ho\v{r}ava--Lifshitz theory in
(3+1)-dimension through $C\left( t\right) /a^{3}$. In Ref. \cite{[2]}, a
realization of the Ho\v{r}ava--Lifshitz gravity as the dynamical
Newton--Cartan geometry was discussed.

The scheme of equations in the projectable version of Ho\v{r}ava--Lifshitz
theory in (3+1)-dimensions is given by the equations

\begin{eqnarray}
\dot{\rho}+3H\left( \rho +p\right) &=&-Q,  \notag \\
3\eta \left( 2\dot{H}+3H^{2}\right) &=&p.
\end{eqnarray}%
where $\eta $ is a dimensionless constant parameter associated with
invariance under diffeomorphisms and $Q$ represents the amount of energy
non-conservation \cite{[3]}. Here there is no Friedmann constraint. From
these equations, it is straightforward to find the first integral

\begin{equation}
3\eta H^{2}=\rho +\frac{C\left( t\right) }{a^{3}}\text{ \ \ }with\text{ \ \
\ }C\left( t\right) =C_{0}+\int_{t_{0}}^{t}dta^{3}Q,
\end{equation}%
and $C\left( t\right) /a^{3}$ is not a real dark matter, but gravitationally
it behaves like a fluid with $p=0$.

Now, we return to the equations given in (\ref{theta2}), which can be
written in the form

\begin{eqnarray}
\dot{\rho}+4H\left( \rho +p\right) &=&0,  \label{4'} \\
\frac{\ddot{a}}{a} &=&\dot{H}+H^{2}=-\frac{3}{8\nu ^{2}}\left( \rho
+2p\right) .  \label{5'}
\end{eqnarray}%
Considering the barotropic relation $p=\omega \rho $, we can write (\ref{4'}%
) and (\ref{5'}) in the form%
\begin{equation}
\dot{\rho}+4H\left( 1+\omega \right) \rho =0,  \label{6'}
\end{equation}%
from which it is direct to obtain%
\begin{equation}
\rho \left( a\right) =\rho \left( a_{0}\right) \left( \frac{a_{0}}{a}\right)
^{4\left( 1+\mathbf{\omega }\right) },
\end{equation}%
and

\begin{equation}
\dot{H}+H^{2}=-\frac{3}{4\nu ^{2}}\left( \omega +\frac{1}{2}\right) \rho ,
\label{7'}
\end{equation}%
from which we can see that $\ddot{a}\gtreqless 0$ and that $\omega
\lesseqqgtr -1/2.$

The equations (\ref{3'}) and (\ref{6'}), with $1+z=a_{0}/a\,$ where $z$ is
the redshift parameter, implies that the Hubble parameter can be written as

\begin{equation}
H\left( z\right) =\sqrt{\frac{3\rho \left( 0\right) }{8\nu ^{2}}\left(
1+z\right) ^{4\left( 1+\mathbf{\omega }\right) }+\left( H^{2}\left( 0\right)
-\frac{3\rho \left( 0\right) }{8\nu ^{2}}\right) \left( 1+z\right) ^{2}}.
\label{8'}
\end{equation}%
We considere now some particular cases:

\begin{enumerate}
\item[(1)] \textbf{Case when }$\mathbf{\omega =0}$. In this case $\rho
\left( z\right) =\rho \left( 0\right) \left( 1+z\right) ^{4}$ and
\end{enumerate}

\begin{equation}
H\left( z\right) =\sqrt{\frac{3\rho \left( 0\right) }{8\nu ^{2}}\left(
1+z\right) ^{2}+\left( H^{2}\left( 0\right) -\frac{3\rho \left( 0\right) }{%
8\nu ^{2}}\right) }\left( 1+z\right) ,  \label{9'}
\end{equation}%
where we can see that%
\begin{equation}
H\left( z\rightarrow \infty \right) \rightarrow \infty \text{ \ \ \ }and%
\text{ \ \ }H\left( z\rightarrow -1\right) \rightarrow 0.  \label{10'}
\end{equation}

\begin{enumerate}
\item[(2)] \textbf{Case when }$\mathbf{\omega =-1/2}$. In this case $\rho
\left( z\right) =\rho \left( 0\right) \left( 1+z\right) ^{2}$ and
\end{enumerate}

\begin{equation}
H\left( z\right) =H\left( 0\right) \left( 1+z\right) \Longrightarrow \ddot{a}%
=0,  \label{11'}
\end{equation}%
which correspond to a Milne universe.

\begin{enumerate}
\item[(3)] \textbf{Case when }$\omega =-1$. In this case $\rho \left(
z\right) =\rho \left( 0\right) =const.$, but according to (\ref{8'})
\end{enumerate}

\begin{equation}
H\left( z\right) =\sqrt{\frac{3\rho \left( 0\right) }{8\nu ^{2}}+\left(
H^{2}\left( 0\right) -\frac{3\rho \left( 0\right) }{8\nu ^{2}}\right) \left(
1+z\right) ^{2}},  \label{12'}
\end{equation}%
from which we see%
\begin{equation}
H\left( z\rightarrow \infty \right) \rightarrow \infty \text{ \ \ \ }and%
\text{ \ \ }H\left( z\rightarrow -1\right) \rightarrow \sqrt{\frac{3\rho
\left( 0\right) }{8\nu ^{2}}},  \label{13'}
\end{equation}%
and unlike to General Relativity in (3+1)-dimensions, we have $H\left(
z\right) \neq const$. for $\rho \left( z\right) =const$.

\begin{enumerate}
\item[(4)] \textbf{Case when }$\omega <-1$. In this case $\rho \left(
z\right) =\rho \left( 0\right) \left( 1+z\right) ^{-4\left( \left\vert 
\mathbf{\omega }\right\vert -1\right) }$ and
\end{enumerate}

\begin{equation}
H\left( z\rightarrow \infty \right) \rightarrow \sqrt{H^{2}\left( 0\right) -%
\frac{3\rho \left( 0\right) }{8\nu ^{2}}}\left( 1+z\right) \rightarrow
\infty ,  \label{14'}
\end{equation}%
\begin{equation}
H\left( z\rightarrow -1\right) \rightarrow \sqrt{\frac{3\rho \left( 0\right) 
}{8\nu ^{2}}}\left( 1+z\right) ^{-2\left( \left\vert \mathbf{\omega }%
\right\vert -1\right) }\rightarrow \infty .  \label{15'}
\end{equation}%
We note that $H\left( z\rightarrow -1\right) $ diverges, when $\rho \left(
z\rightarrow -1\right) $. However this not happens in a finite time how it
happens in General Relativity in $(3+1$)-dimensions when we think, for
instance, in a Little Big Rip future singularity \cite{[4]}.

Now, the first and second law of thermodynamics tell us, respectively,

\begin{eqnarray}
TdS &=&d\left( \rho V\right) +pdV=\left( \rho +p\right) dV+Vd\rho ,
\label{16'} \\
T\frac{dS}{dt} &=&V\left[ \dot{\rho}+4H\left( \rho +p\right) \right] .
\label{17'}
\end{eqnarray}%
Since, according to (\ref{4'}), $\dot{\rho}+4H\left( \rho +p\right) =0,$ we
have an adiabatic evolution. This means that in Newton--Chern--Simons
cosmology there is no Friedmann constraint and we have adiabatic evolution.
On the other hand in Ho\v{r}ava--Lifshitz theory in (3+1)-dimensions there
is no Friedmann constraint and, unlike of Newton--Chern--Simons cosmology,
the evolution is non-adiabatic since $\dot{\rho}+4H\left( \rho +p\right) =-Q$
$\neq 0$ and therefore $dS/dt\neq 0$.

In summary we can say that we have presented cosmological schemes from the
five-dimensional Einstein-Chern-Simons gravity theory. It could be
interesting to use some process of compactification to project these results
to (3 + 1) -dimensions and then compare them with the results obtained in
the context of general relativity.

\section{\textbf{Final Remarks}}

We have considered a five-dimensional action $S=\int L_{\mathrm{ChS}%
}^{(5)}+\kappa L_{M}$ which is composed of a gravitational sector and a
matter sector, where the gravitational sector is given by a
Newton--Chern--Simons gravity action instead of the Einstein--Hilbert action
and the matter sector is described by a perfect fluid. We have studied the
implications of replacing the Einstein--Hilbert action by the
Newton--Chern--Simons action on the cosmological evolution for a FLRW metric.

We have showed that the Newton--Chern--Simons cosmology is a sort of
analogue of the projectable version of the Ho\v{r}ava--Lifshitz theory in
(3+1)-dimension, although a term that contains $Q$ is not present. We have
found solutions and their asymptotic limits which show interesting
properties. In addition, a phantom solution with a future singularity
reminiscent of a Litlle Big Rip future singularity has been obtained.
Finally, a brief revision of the adiabaticity in the cosmic evolution was
made.

As we said at the end of the previous section, an interesting thing would be
to do a compactification of five to four-dimensions in order to obtain
generalized non-relativistic cosmologies to be compared with the respectives
schemes studied in the context of general relativity (work in progress).

\textbf{Acknowledgement.} \ \textit{This work was supported in part }by 
\textit{\ FONDECYT Grant  }No.\textit{\ 1180681} from the Government of
Chile. \textit{One of the authors (}GR\textit{) was supported by grant from
Comisi\'{o}n Nacional de Investigaci\'{o}n Cient\'{\i}fica y Tecnol\'{o}gica 
}CONICYT\textit{\ }No.\textit{\ }21140971\textit{\ and from Universidad de
Concepci\'{o}n, Chile. }

\end{document}